\documentstyle [12pt,epsf]{article}
\begin{document}
\begin{titlepage}
\noindent
{\it Phys. Rev. D{\bf 54}, 10/1/96}
\hspace{3.2in}
{\it hep-ph/9608252}
\vspace{0.5in}
\begin{Large}
\begin{center}
{\bf Pion Helicity Structure And Its Consequence For The Hard 
Scattering Form Factor}
\end{center}
\end{Large}
\vspace{5 mm}
\begin{large}
\begin{center}
{S.W. Wang\footnote{Present address: Department of Physics, University 
of Maryland, College Park, MD 20742} and L.S. Kisslinger}
\end{center}
\end{large}
\vspace{5 mm}
\begin{center}
{Physics Department, Carnegie-Mellon University, Pittsburgh, 
PA 15213, USA }
\end{center}
\vspace{1.2in}
\begin{abstract}
In a study of the elastic pion form factor for large momentum
transfers based on a modified perturbative QCD (PQCD) approach we 
have included helicity components that are customarily neglected.
Along with the inclusion of transverse momentum, this gives a large
suppression of the form factor from the prediction 
of the original hard scattering model
based on  PQCD in the $Q^{2}$ domain where experimental data are 
available. We conclude that nonperturbative contributions will dominate 
in this region.
\end{abstract}
\vspace{0.5in}
\begin{center}
\end{center}
\end{titlepage}

It has been suggested that the hard scattering model for the hadronic 
form factors\cite{br} derived using perturbative QCD 
(PQCD) is applicable in the currently experimentally accessible regime
of momentum transfer
($Q^{2}\sim$ a few GeV$^{2}$). Facing critical questions and doubts
\cite{is} about whether it is justified to use the quark distribution 
function proposed by Chernyak-Zhitnitsky (CZ)\cite{cz}, which can
increase the hard scattering form factor to improve consistency with
experiment, Li and Sterman\cite{li} 
recently proposed a scheme with Sudakov suppression 
included in an attempt to make the hard scattering formalism internally
consistent. They started with a modified
expression of hard scattering amplitude $T_{H}$ in which they had 
retained the transverse momentum in the gluon propagator.
The application of the renormalization group
to both the wave function and the hard scattering amplitude $T_{H}$ 
allowed them to pinpoint the dangerous soft end-point region and associate 
it with the transverse distance. They conclude that with the 
inclusion of the Sudakov correction the hard scattering form remains 
valid and [with a CZ quark distribution] is comparable to the experimental 
data for the $Q^{2}$ in as low as about 1 GeV$^{2}$ for the pion. 

However, in our early work on the calculation of pion 
elastic form factor in the light-cone Bethe-Salpeter (BS) 
formalism\cite{kis}, and in our recent work\cite{kw}
in which a Sudakov form factor and other effects have been included,
we consistently find a dominant soft contribution (i.e. form factor arising 
from the confining kernel as defined in\cite{kis,kw}) in the few GeV$^{2}$
region. With the inclusion of transverse 
momentum in both the soft wave function and the one-gluon exchange integral
kernel there is strong 
suppression of hard scattering at a few GeV$^{2}$.  This effect has also 
been pointed out by Jacob and Kroll in 
their recent work\cite{jak}, but they have dropped helicity components
in the soft wave function which do not contribute in the conventional model. 

With the inclusion of transverse momentum there are contributions
to the hard scattering process which are not included in the 
standard\cite{br} hard scattering formula.
It is the objective of the present note to reanalyse the pure hard 
scattering pion form factor using the modified $T_{H}$ but without the 
exclusion of the transverse momentum  or the unconventional helicity 
components in the soft wave function. 
We show that these effects produce important changes in the results of
the hard scattering prediction in the region of momentum transfer
where current experimental data are available. In this note 
we will be using two models for the soft wave function which give 
satisfactory  descriptions for low $Q^{2}$ properties of the form factor. 

Let us first review the light-cone wave functions [BS amplitudes]
needed for the hard scattering form.
The transformation of the single particle states between different 
representations can be formulated using their identity at zero 
momentum. For instance, the instant form of a single-particle state is
related to the light-cone representation by the following equation\cite{kt}, 

\begin{eqnarray}
\mid {\bf p}\mu \rangle & = & \sum_{\bar{\mu}}
\mid {\bf p}\bar{\mu} \rangle_{l.c.} D^{j}_{\bar{\mu}\mu}[R_{fc}(p)],
\label{eq:melosh}
\end{eqnarray}
where $R_{fc}(p)$ is called the Melosh rotation
\cite{mel} and the subscript $l.c$ denotes light-cone representation.
It is easy to work out Eq. (\ref{eq:melosh}) for $j=1/2$, which is\cite{kt}
 
\begin{equation}
\mid {\bf p}\uparrow (\downarrow) \rangle =
\frac{(p^{+}+m)\mid {\bf p}\uparrow (\downarrow) \rangle_{l.c.} \mp 
p^{R(L)} \mid {\bf p} \downarrow (\uparrow) \rangle _{l.c.}}
{\sqrt{2p^{+}(p^{0}+m)}},
\label{eq:state}
\end{equation}
where $p^{R(L)}=p_{x}\pm ip_{y}$ and $m$ is the quark mass.

This result, that a helicity state in an instant-form representation is
a mixed helicity state in the light-cone representation,
is hardly a surprise since a ``rotationless boost'' defined in 
the light-cone representation is not a pure boost any more when viewed in 
the instant form and vice versa. In other words, the transform from one
representation to another inevitably causes a state to undergo a 
``rotation''.  

To construct a state for a pion in the light-cone form, one starts with a 
quark-antiquark pair being coupled to form a spin zero state in the pion 
rest frame, and then boosts it to an arbitrary momentum  using
Eq. (\ref{eq:state}).  Using the standard light-cone notation of
$(x,{\bf k_{\bot}})$ for the longitudinal and transverse relative momentum, 
we tabulate the explicit form of the wave functions along with
their Fourier transforms in the Table 1 and Table 2. One sees that there
do exist four helicity components, namely

\begin{equation}
\Psi=(\Psi_{\uparrow \uparrow},\Psi_{\uparrow \downarrow}
,\Psi_{\downarrow \uparrow},\Psi_{\downarrow \downarrow}).
\label{eq:pispin}
\end{equation}
  
For simplicity of illustration, in the present work we consider
two models of pion helicity functions. One is obtained by a naive extension 
from nonrelativistic spin form to the relativistic light-cone form 
\begin{equation}
\frac{1}{\sqrt{2}}\chi^{\dagger}(1)i\sigma_{2} \chi (2) \longrightarrow 
\chi_{\pi}=\frac{1}{\sqrt{2}}\bar{u}(1)\gamma_{5}v(2).
\label{eq:naive}
\end{equation}
Another is suggested by Dziembowski\cite{dz}
\begin{equation}
\chi_{\pi}=\bar{u}(1)(p_{\mu}\gamma^{\mu}+m_{\pi})\gamma_{5}v(2).
\label{eq:dzb}
\end{equation}
The light-cone spinors $u$ and $v$ have the Melosh rotation built into 
them. The helicity function given by Eq. (\ref{eq:dzb}) is
a four-component generalization of the one obtained by a Melosh rotation
of the wave function given on the left-hand side of Eq. (\ref{eq:naive}),
apart from an overall factor. 
For the confinement wave functions in momentum space we have adapted
the Brodsky-Huang-Lepage's oscillator prescription\cite{br}. 

The modified $T_{H}$ is obtained by retaining the transverse momentum in 
the gluon propagator\footnote{The transverse momentum in the quark
propagator has been neglected for the simplicity of analysis. This is
sufficient for the purpose of our present work for it is known \cite{li}
that such effect gives about 15\% 
correction for the form factor at 2 GeV$^{2}$, about an order of magnitude
smaller than the effects we find for additional helicity components.
Moreover, these additional corrections would further suppress the PQCD
contribution.}, which gives
\begin{equation}
T_{H}(x,y,{\bf k_{\bot}},{\bf \ell_{\bot}},Q,\mu)=\frac{4g^{2}(\mu)C_{F}}
{(1-x)(1-y)Q^{2}+({\bf k_{\bot}}-{\bf \ell_{\bot}})^{2}}.
\label{eq:th}
\end{equation}
Li-Sterman\cite{li} introduced a Fourier transform in transverse momentum,
leading to the modified hard scattering pion form factor:

\begin{equation}
F_{\pi}(Q^{2})=\int \frac{dxdy}{(16\pi^{3})^{2}} \int \frac{d^{2}b}
{(2\pi)^{2}}\hat{\Psi}^{\ast}(y,{\bf b})\hat{T}_{H}(x,y,{\bf b},Q,t)
\hat{\Psi}(x,{\bf b})\exp(-S(x,y,b,Q,t)),
\label{eq:ff}
\end{equation}
where the Fourier transform of $T_{H}(x,y,{\bf k_{\bot}},
{\bf \ell_{\bot}},Q,\mu)$ is
\begin{equation}
\hat{T}_{H}(x,y,{\bf b},Q,t)=32\pi^{2}C_{f}\alpha_{s}(t)K_{0}
(\sqrt{(1-x)(1-y)}Qb),
\label{eq:thb}
\end{equation}
with $K_{0}(z)$ being the modified Bessel function, and 

\begin{equation}
t=\max(\sqrt{(1-x)(1-y)}Q,1/b).
\label{eq:tmax}
\end{equation}
The Sudakov factor $S(x,y,b,Q,t)$ is given in detail in\cite{li}.
In deriving Eq. (\ref{eq:th}) and Eq. (\ref{eq:ff}), it has been assumed 
that only those conventional helicity components, namely $h_{1}+h_{2}=0$,
are relevant. In order to include the $h_{1}+h_{2}=\pm1$ components in 
addition, one needs to find out their corresponding $T_{H}$.

Taking into account the symmetries of our model wave functions, especially 
the fact that $\Psi^{\ast}_{\uparrow \uparrow}(x,{\bf k_{\bot}})= 
\Psi_{\downarrow \downarrow}(x,{\bf k_{\bot}})$, and under the assumption 
that $T_{H}$ depends only on ${\bf k}={\bf k_{\bot}}-{\bf \ell_{\bot}}$,  
it is straightforward to show that

\begin{equation}
T_{H}(x,y,{\bf k},Q,\mu)^{\uparrow \downarrow +\downarrow \uparrow}
=-T_{H}(x,y,{\bf k},Q,\mu)^{\uparrow \uparrow +\downarrow \downarrow}.
\label{eq:proof}
\end{equation}
One will see later that the  minus sign in $T_{H}$ associated with $\uparrow 
\uparrow +\downarrow \downarrow$ components here has the crucial consequence 
that the unconventional helicity components will indeed generate a 
suppression on the pion form factor. 
 
\begin{table}[hb]
\begin{center}
\caption{The naive wave function $\Psi(x,{\bf k_{\bot}})$ and its 
FT $\hat{\Psi}(x,{\bf b})$} 
\vspace{0.2cm}
\begin{tabular}{|c|c|c|}
\hline
\multicolumn{1}{|c|}{$\lambda_{1} \lambda_{2}$} 
& \multicolumn{1}{c|}
{$\Psi_{\lambda_{1}\lambda_{2}}(x,{\bf k_{\bot}})$} 
& \multicolumn{1}{c|}
{$\hat{\Psi}_{\lambda_{1}\lambda_{2}}(x,{\bf b})$}  \\ \hline
$\uparrow \uparrow$ 
& -$k^{L}\hat{\phi}(x,k_{\bot}) $
& $ib^{L}\xi(x,b)g(x)/2 $ \\ \hline
$\uparrow \downarrow$
& $m\hat{\phi}(x, k_{\bot})$
& $m\xi(x,b)$ \\ \hline
$\downarrow \uparrow$
& -$m\hat{\phi}(x,k_{\bot}) $
& -$m\xi(x,b)$  \\ \hline
$\downarrow \downarrow$
& -$k^{R}\hat{\phi}(x,k_{\bot})$ 
& $ib^{R}\xi(x, b)g(x)/2 $ \\ \hline
\end{tabular}
\end{center}
\end{table}
The complete wave functions and their corresponding Fourier transforms (FT)
are given explicitly in the Table 1 and Table 2, where the FT is defined as
\begin{equation}
\Psi(x,{\bf k_{\bot}})=\int \frac{d^{2}b}{(2\pi)^{2}}\hat{\Psi}(x,
{\bf b})e^{-i{\bf b \cdot k_{\bot}}}.
\label{eq:ft}
\end{equation}

\newpage
\begin{table}[th]
\begin{center}
\caption{The Dziembowski wave function $\Psi(x,{\bf k_{\bot}})$ and its 
FT $\hat{\Psi}(x,{\bf b})$} 
\vspace{0.4cm}
\begin{tabular}{|c|c|c|}
\hline
\multicolumn{1}{|c|}{$\lambda_{1} \lambda_{2}$} 
& \multicolumn{1}{c|}
{$\Psi_{\lambda_{1}\lambda_{2}}(x,{\bf k_{\bot}})$} 
& \multicolumn{1}{c|}
{$\hat{\Psi}_{\lambda_{1}\lambda_{2}}(x,{\bf b})$}  \\ \hline
$\uparrow \uparrow$ 
& -$(a_{1x}+a_{2x})k^{L}\hat{\phi}(x,k_{\bot}) $
& $i(a_{1x}+a_{2x})b^{L}\xi(x,b)g(x)/2 $ \\ \hline
$\uparrow \downarrow$
& $(a_{1x}a_{2x}-k^{2}_{\bot})\hat{\phi}(x, k_{\bot})$
& $(a_{1x}a_{2x}-g(x)(1-b^{2}g(x)/4))\xi(x,b)$ \\ \hline
$\downarrow \uparrow$
& -$(a_{1x}a_{2x}-k^{2}_{\bot})\hat{\phi}(x,k_{\bot})$ 
& -$(a_{1x}a_{2x}-g(x)(1-b^{2}g(x)/4))\xi(x,b)$  \\ \hline
$\downarrow \downarrow$
& -$(a_{1x}+a_{2x})k^{R}\hat{\phi}(x,k_{\bot}) $
& $i(a_{1x}+a_{2x})b^{R}\xi(x, b)g(x)/2 $ \\ \hline
\end{tabular}
\end{center}
\end{table}

The notation used in Tables 1 and 2 is: 
\begin{eqnarray*}
& a_{ix} & =x_{i}m_{\pi}+m,\;\; \hat{\phi}(x, k_{\bot})=\frac{
\exp(-\frac{k^{2}_{\bot}+m^{2}}{8\beta^{2}x(1-x)})}{x(1-x)},\; \; 
k^{R(L)}=k_{x}\pm ik_{y}, \\
&b^{R(L)} & =b_{x}\pm ib_{y},\; \; g(x)=8\beta^{2}x(1-x), \; \; 
\xi (x,b)=8\pi\beta^{2}\exp(-\frac{b^{2}g(x)}{4}-\frac{m^{2}}{g(x)}). 
\end{eqnarray*}

Writing Eq. (\ref{eq:ff}) in terms of helicity components, we get
\begin{eqnarray}
F_{\pi}(Q^{2})& = &\int \frac{dxdy}{(16\pi^{3})^{2}} \int \frac{bdb}
{2\pi}\hat{T}_{H}(x,y,{\bf b},Q,t)
[\hat{\Psi}^{\ast}_{\uparrow \downarrow}(y,{\bf b})
\hat{\Psi}_{\uparrow \downarrow}(x,{\bf
b})+\hat{\Psi}^{\ast}_{\downarrow \uparrow}(y,{\bf b})
\hat{\Psi}_{\downarrow \uparrow}(x,{\bf b})  \nonumber \\
	      &   &-(\hat{\Psi}^{\ast}_{\uparrow \uparrow}(y,{\bf b})
\hat{\Psi}_{\uparrow \uparrow}(x,{\bf b})+\hat{\Psi}^{\ast}_{\downarrow 
\downarrow}(y,{\bf b})\hat{\Psi}_{\downarrow \downarrow}(x,{\bf b}))]
\exp(-S(x,y,b,Q,t)).
\label{eq:new}
\end{eqnarray}

Before carrying out the calculation, we have to fix the parameters in the 
wave function. The constraint from $\pi \longrightarrow \mu \bar{\nu}$
decay requires that the wavefunction be normalized by the pion decay
constant, $f_{\pi}$, through the relation
\begin{equation}
\int \frac {dxd^{2}k_{\bot}}{16\pi^{3}}\Psi(x,{\bf k_{\bot}})=\frac
{f_{\pi}}{2\sqrt{n_{c}}}, 
\label{eq:fpi}
\end{equation}
where $f_{\pi}=$93  MeV and $n_{c}=3$. In order for the theory to be 
physically sensible and self consistent, we make sure that the parameters, 
i.e., quark mass $m$ and harmonic oscillator parameter $\beta$, are chosen 
in such a way that while the equation (\ref{eq:fpi}) is satisfied, the
calculated  pion charge radius is approximately equal to its experimental
value  ($<\!r_{\pi}^{2}\!>^{1/2}_{\exp}= 0.66 \: fm$\cite{da}).

\begin{figure}[th]
\hskip 1in
\epsfxsize=4.66in
\epsfysize=2.09in
\epsffile{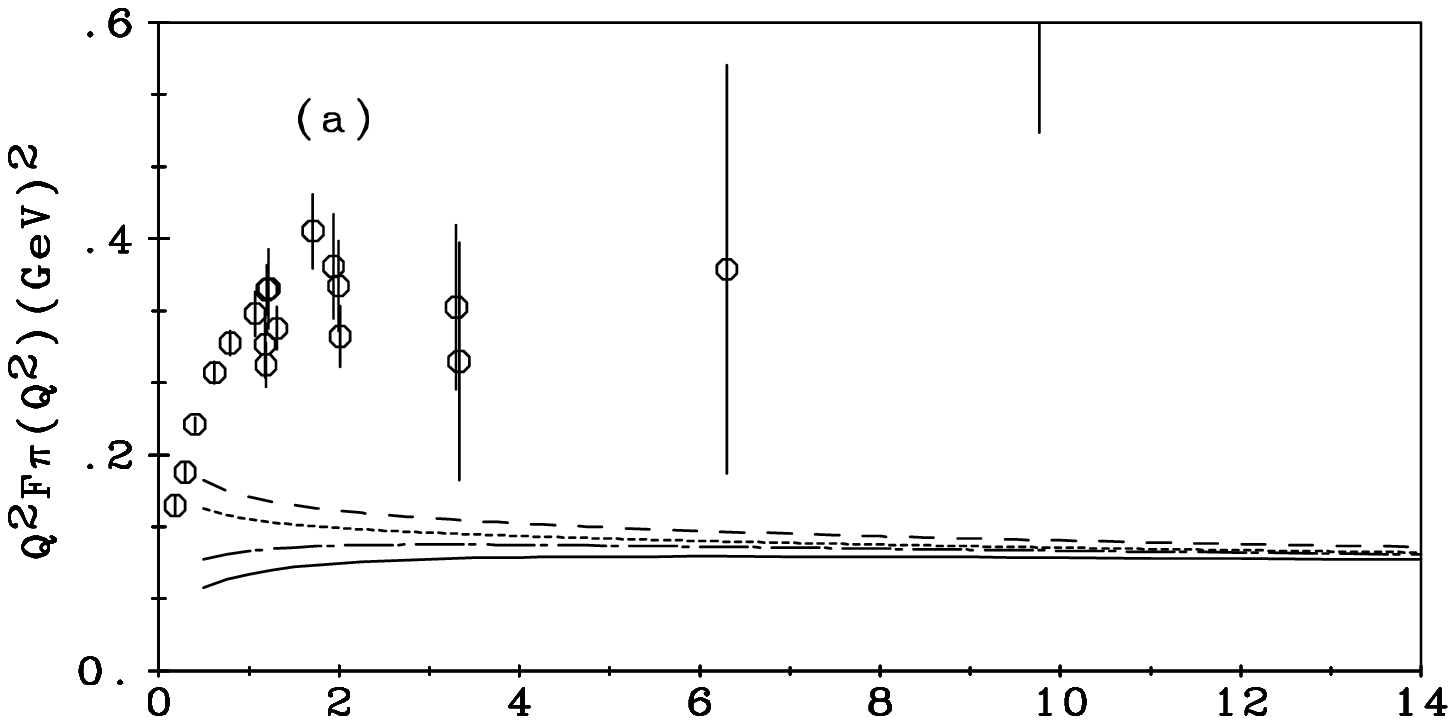}
\vskip 0.1in
\hskip 1in
\epsfxsize=4.66in
\epsfysize=2.09in
\epsffile{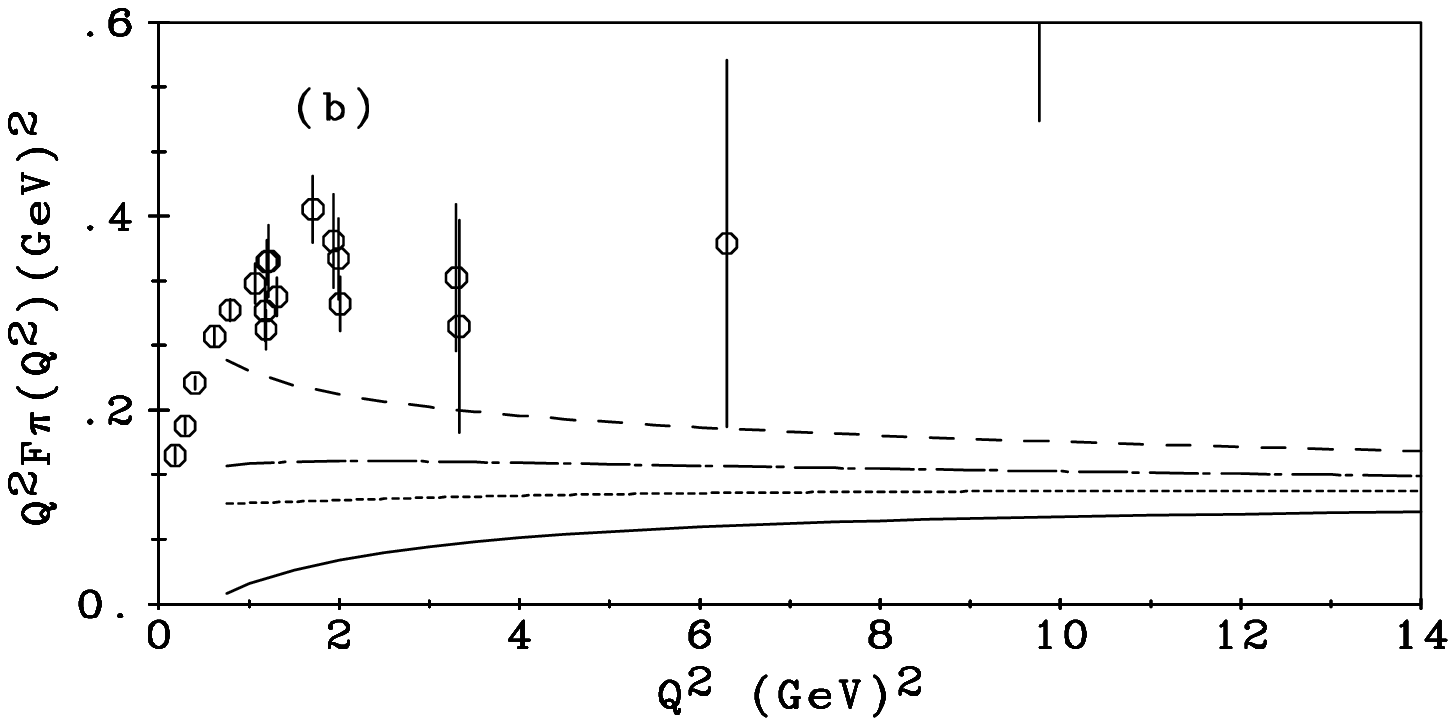}
\caption{Pion form factors. The dashed and dotted lines are the results 
without or with $h_{1}+h_{2}=\pm1$ components, respectively. The 
dash-dotted and solid lines are the corresponding results but with the 
Sudakov correction included. The naive helicity function Eq. (\protect
\ref{eq:naive}) is used for Fig. ($a$) with $\beta \!=\!260$ MeV, while 
for Fig. ($b$), Dziemboski's helicity function Eq. (\protect\ref{eq:dzb}) 
is used with  $\beta \!=\!320$ MeV. 
The parameters: $m\!=\!330$ MeV and $\Lambda_{QCD}=200$ MeV are the
same for both ($a$) and ($b$). The experimental data are taken from\protect
\cite{be}.}
\label{fig_1} 
\end{figure}

The results of our numerical calculation are presented in Fig. \ref{fig_1}, 
where the difference in
$Q^{2}F_{\pi}(Q^{2})$ with the inclusion and the exclusion of 
$h_{1}+h_{2}=\pm1$ components as well as the Sudakov effects are shown. 
The curves (a) and (b) are for the naive helicity function and Dziembowski's 
form, respectively. 

There are several aspects worth commenting here with respect to our 
calculation: 1) It is necessary to satisfy the physical constraints at low 
$Q^{2}$ and Eq. (\ref{eq:fpi}) if one is to study the transition from soft to
hard QCD.  With $m=330$ MeV, $\beta=260$ (320) MeV for model 1 (2), we get
$<r_{\pi}^{2}>^{1/2} =$0.65 (0.64) fm, respectively. It has been known that 
with a specific set of 
\noindent
parameters, one can obtain a quark distribution amplitude (QDA) similar 
to that of CZ's from Dziembowski's wave function. However, using
this set of parameters we have failed to produce any hard scattering form 
factors which are physically sensible below $Q^{2}=100$ GeV$^{2}$.
Other problems with this set of parameters have been pointed out in\cite{ji}. 
This again shows that the $k_{\bot}$ dependence is not as trivial as one might 
have naively speculated. 2) The numerical results with only $h_{1}+h_{2}=0$ 
components are similar to the standard hard scattering prediction using the 
asymptotic form of the QDA, which is likely due to the fact that 
the QDAs obtained from our wave functions are close in form to the 
asymptotic one. 3) Adding the $h_{1}+h_{2}=\pm1$ contents suppresses 
the hard scattering significantly in the present experimentally accessible 
energy region regardless the Sudakov correction. This suppression shows
up for both helicity model wave functions but much stronger with 
Dziembowski's form as can be seen from Fig. 1($b$).

4) The contribution from $h_{1}+h_{2}=\pm1$ becomes less important in
large $Q^{2}$. For model 1, this contribution vanishes beyond 
$Q^{2}=$20  GeV$^{2}$. For model 2, the contribution is less than 5\% 
at about 100 GeV$^{2}$ while it generates a suppression of factor two 
for the form factor at 2 GeV$^{2}$. As $Q^{2}$ goes to infinity, we recover 
the asymptotic prediction for both models. Therefore the conventional hard 
scattering theory remains intact only when $Q^{2}\rightarrow \infty$. 5) 
Finally, in our analysis we neglect the quark propagator in the $T_{H}$
for our focus is on the helicity structure itself. This is justified by
our result that the suppression from the unconventional helicity components 
is nearly 10 time larger than the effect from inclusion of the quark 
propagator at 2 GeV$^{2}$ for the model wavefunction 2.

In conclusion, we have re-investigated the hard
scattering pion form factor using the modified scattering amplitude 
$T_{H}$ with transverse momentum included. 
Special attention has been paid to the unconventional helicity
components which are customarily neglected. We have shown that inclusion
of these components suppresses the hard scattering at moderate
$Q^{2}$ due to the fact that $T_{H}$ associated with $h_{1}+h_{2}=\pm1$ 
bears an opposite sign in comparison with $h_{1}+h_{2}=0$ portions.
Although this analysis is carried out using two particular 
models, the general consistency of the picture is certainly a further 
indication that the hard scattering mechanism is not the dominant one
in the moderate $Q^{2}$ region, where experimental results are available.  
Therefore, there must be a quite sizable soft contribution as we have 
pointed  out in\cite{kis,kw}. 

{\em Acknowledgement}: This work is supported in part by NSF grant
PHY-9023586 and in part by DOE Grant DOE-FG02-93ER-40762.

\end{document}